\documentclass[reprint,
superscriptaddress,
 amsmath,amssymb,
 aps,
pra,
]{revtex4-2}

\usepackage{graphicx}
\usepackage{dcolumn}
\usepackage{xcolor}
\usepackage{bm}
\usepackage{physics}
\usepackage{hyperref}
\hypersetup{
	colorlinks = true,
	urlcolor   = blue,
	linkcolor  = blue,
	citecolor  = red
}

\begin{document}

\title{Efficient lineshape estimation by ghost spectroscopy}
\author{Ilaria~Gianani}
\affiliation{Dipartimento di Scienze, Universit\`a degli Studi Roma Tre, Via della Vasca Navale~84, 00146 Rome, Italy} 
\email{ilaria.gianani@uniroma3.it}
\author{Luis L. S\'anchez-Soto}
\affiliation{Departamento de \'{O}ptica, Facultad de F\'{\i}sica, Universidad Complutense, 28040 Madrid, Spain}
\affiliation{Max-Planck-Institut f\"{u}r die Physik des Lichts, 91058 Erlangen, Germany}
\author{Aaron Z. Goldberg}
\affiliation{National Research Council of Canada, Ottawa, Ontario K1N 5A2, Canada}
\affiliation{Department of Physics, University of Ottawa, Ottawa, Ontario K1N 6N5, Canada}
\author{Marco~Barbieri}
\affiliation{Dipartimento di Scienze, Universit\`a degli Studi Roma Tre, Via della Vasca Navale~84, 00146 Rome, Italy}
\affiliation{Istituto Nazionale di Ottica - CNR, Largo Enrico Fermi 6, 50125 Florence, Italy}

\begin{abstract}
Recovering the original spectral lineshapes from data obtained by instruments with extended transmission profiles is a basic tenet in spectroscopy. By using the moments of the measured lines as basic variables, we turn the problem into a linear inversion. However, when only a finite number of these moments are relevant, the rest of them act as nuisance parameters. These can be taken into account with a semiparametric model, which allows us to establish the ultimate bounds on the precision attainable in the estimation of the moments of interest. We experimentally confirm these limits with a simple ghost spectroscopy demonstration.   
\end{abstract}

\maketitle

Spectroscopy is one of the most relevant methods to explore the structure of matter and has played a major role in various fields of science and technology. In general, obtaining spectral information involves two different steps: data import (preprocessing) and information extraction (postprocessing). In the first stage, raw data is transformed to \emph{cleaned} data to remove artifacts caused during measurement and thus improve prediction performance~\cite{Rinnan:2009aa}. Postprocessing identifies and quantifies the spectral lines, which requires resolving overlapped spectral components. 

Overlapping of spectral components is due to their broadening, possibly caused by instrumental distortions.  Decomposing the spectrum into individual lines thus requires the use of spectral deconvolution, an inverse problem commonly encountered in signal processing. The topic has been the subject of numerous studies. Derivative spectroscopy, usually restricted to the first- and second-order, was one of the first deconvolution methods and it is still widely used~\cite{Dubrovkin:2021aa}. Fourier self-deconvolution is another popular option~\cite{Kauppinen:1981aa}. Effective deconvolution algorithms based on sophisticated mathematical tools, such as digital filters~\cite{Gelb:1974aa,Sprzeczak:2001aa}, spline approximations~\cite{Averbuch:2009aa}, and Tikhonov regularization~\cite{Liu:2013aa}, can be found in the literature under a variety of names. More recently, machine learning techniques, such as neural networks~\cite{Ronneberger:2015aa,Yanny:2022aa} and random forest~\cite{Torrisi:2020aa},  have been applied with notable success to this problem. Accordingly, many useful software tools for spectral deconvolution are available to the practitioner in the field.

While these approaches are efficient for the practical reconstruction of the spectral features, they do not provide any hint as to the ultimate  precision limits attainable in any experiment. It seems thus pertinent to revisit this problem from a modern quantum viewpoint. This not only will allow us to ascertain those tight bounds, but it can identify optimal measurement strategies, as has been recently demonstrated for some related problems~\cite{Albarelli:2020aa,Polino:2020aa}. Here we present the application of semiparametric estimation to the problem of spectral deconvolution in the context of ghost spectroscopy, which has emerged as an intriguing possibility for remote sensing~\cite{Amiot:2018aa,Janassek:2018aa,Chiuri:2022aa}. Taking inspiration from imaging~\cite{Tsang:2019aa,Cimini:2021aa}, we can use semiparametric estimation to derive proper bounds on the precision without relying on large matrix inversion and quantify how the uncertainties on the instrumental function of our spectrometer impact our estimations. 

\begin{figure}[t!]
\includegraphics[width=\columnwidth]{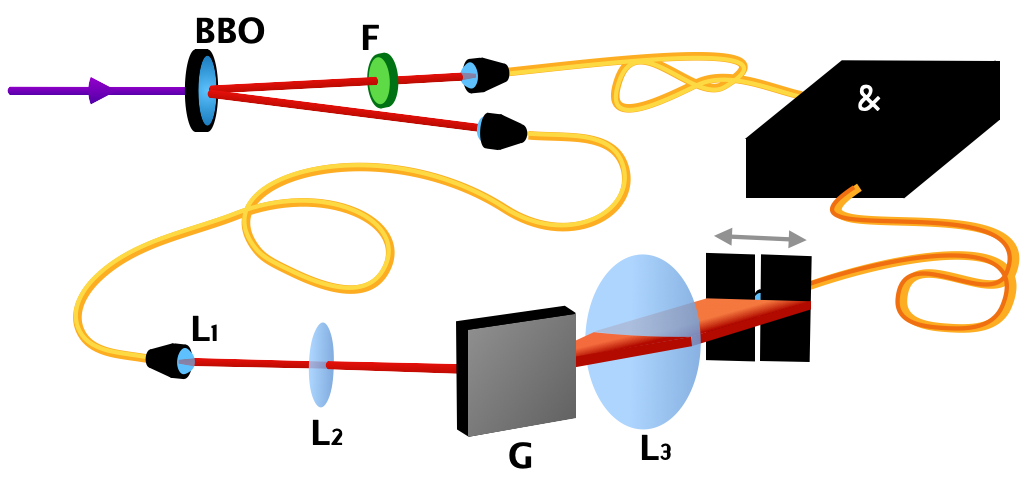}
\caption{Experimental setup. A 30-mW CW diode laser at 405~nm pumps a 3-mm barium borate (BBO) crystal cut for noncollinear Type-I phase matching. One photon is sent through the spectral absorption object and detected with a bucket detector. Frequency detection is performed on the second photon: a telescope, composed of a collimator $L_1$ and an $f=150$ mm lens $L_2$, collimates the beam on a 1200-lines/mm grating, which is then focused by a lens $L_3$ with focal length $f = 50$~mm on a 100-$\mu$m slit. Finally, a multimode fibre collects the signals for a complete measurement of the spectral range under investigation by raster scanning.}
\label{fig:setup}
\end{figure}

For our investigation, we have employed the arrangement illustrated in Fig.~\ref{fig:setup}. We generate photon pairs by parametric downconversion (PDC); this provides a correlated pair in the time-frequency domain, serving as the basis for remote spectral profile reconstruction. The working principle is that spectral information about the absorption of a sample can be retrieved by leveraging such frequency correlations~\cite{Amiot:2018aa}, even when the arm with the sample is not monitored with a frequency-sensitive detector. The idler photon reaches the spectral object and is then measured by a frequency-insensitive ``bucket" detector. The signal photon is spatially filtered using a single-mode fibre as well, and is then sent to a home-built spectrometer. Because of the frequency correlation, the level of the coincidence counts will be proportional to the transmittance at that photon energy.

\begin{figure}[t!]
\includegraphics[width=\columnwidth]{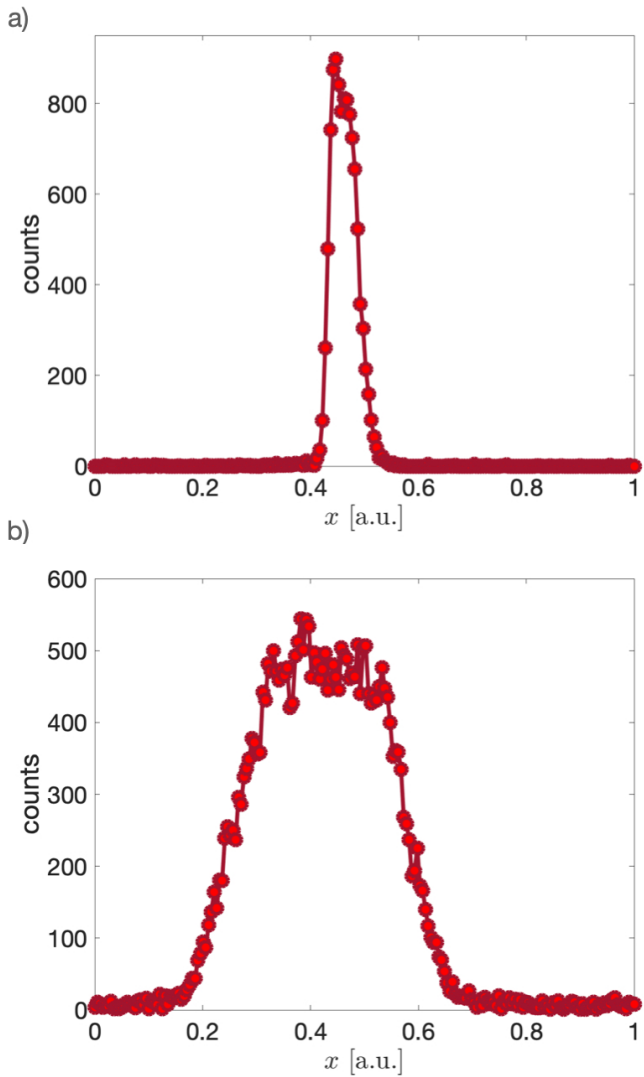}
\caption{a) Experimentally determined profile of the instrumental function $H(x)$:  in our experiment, we adopted a filter centred at 853~nm with a full-width at half-maximum of 0.85~nm. b) Experimental profile of the spectral object: this is a supergaussian filter with centre wavelength  810~nm and a full-width at half-maximum of 7.3~nm. The $x$-axis is in arbitrary units normalised to the useful scan range.}
\label{fig:profiles}
\end{figure}

Inferring the properties of the true spectrum $F(x)$ from the measured spectrum $f(x)$ requires solving the Fredholm first-kind integral equation
\begin{equation}
\label{eq:conv}
f(x) = \int H(x- x^{\prime}) F(x^{\prime}) \, dx^{\prime} \, ,
\end{equation}
where $H(x)$ is the instrumental function (also known, depending on the context, as the impulse response or point spread function, see Supplementary Material).  The smoothing properties of the kernel $H(x)$ are responsible for the information loss in the convolution process~\cite{Pike:1984aa}.  Additionally, quite a few mathematical models of instrumental functions are at hand~\cite{Rautian:1958aa}. In our experiment, this can be retrieved by placing a narrowband filter in the idler arm, whose width is narrower than the resolution of the spectrometer, leading to the results in Fig.~\ref{fig:profiles}a. This is akin to what is the standard in astronomical observations, where a distant star serves as a reference to obtain the point spread function of the imaging system~\cite{Chen:2021aa}.

An interferometric filter serves as our spectral object: an example a reconstructed profile appears in Fig.~\ref{fig:profiles}b. We should notice that the two filters cover different spectral regions: space-time coupling occurs in PDC, thus the employ of single-mode fibres, although beneficial to the signal-to-noise ratio, can limit the available frequency range for a given phase-matching configuration. We have solved this issue by optimising the spatial coupling for each measurement, so as to ensure that the phase-matching conditions are the same when carrying out the measurement for the instrumental function and those of the spectral profiles.

In the framework of metrology, the measured spectrum is regarded as a collection of random variables, and it is typically assumed that these are described by a parametric statistical model; i.e., a known dependence of the lineshape on a finite number of parameters. Then, an estimator attempts to predict the unknown parameters solely using the measurements. The time-honored Cram\'er-Rao bound then gives the optimal precision by inverting the Fisher information matrix~\cite{Kay:1993aa}. However, when the data cannot be fitted \emph{a priori}  to a certain probability distribution, the model becomes infinite-dimensional. One would need to evaluate and invert the Fisher information matrix for all of the parameters present in the model, even those whose values will never be estimated.  This is prohibitive, to say the least, and may be circumvented by turning to tools recently forged for the theory of quantum metrology. To this end, we consider as our parameters of interest the moments of the true line profile
\begin{equation}
\label{parameters}
M_{i} = \int x^i F(x) \, dx.
\end{equation}
These moments (that we assume to be finite valued) have been considered profitably in several contexts.  They have been evaluated for the absorption of light by atoms and molecules~\cite{Gordon:1965aa}; they have been used  to study the potentials involved in collision-induced absorption~\cite{Leine:1968aa}, in lineshape theories~\cite{Jacobson:1971aa} and to infer intermolecular-force information~\cite{Jacobson:1971ab}.

These parameters, however, are not directly accessible from the measured profile $f(x)$. Instead, we have to consider its moments
\begin{equation}
m_{i} = \int x^i f(x) \, dx \, .
\end{equation}
The crucial observation for what follows is that these moments are related through the lower triangular matrix formed by the moments of the instrumental function, which we denote by $\mathbf{C}$, with elements
\begin{equation}
    {C}_{ij}= \binom{i}{j}\int H(x) x^{i-j} dx.
\end{equation}
As a consequence, the inversion of \eqref{eq:conv} reduces to the infinite hierarchy~\cite{Tsang:2019aa} 
\begin{equation}
  M_{i} = \sum_{j} (\mathbf{C}^{-1})_{ij} \; m_{j} \,,
\end{equation}
with $i= 0,1, 2, \ldots$. This is instrumental to constructing estimators for the $M_i$'s by starting from the estimation of the $m_i$'s. This will unavoidably consider a set of discrete data points taken at positions $x_k$ approximating the actual shape of $f(x)$, thus leading to an estimator:
\begin{equation}
\label{estimator}
    \check{m}_i = \frac{1}{N}\sum_k (x_k)^i N(x_k),
\end{equation}
where $N(x_k)$ is the count rate at the position $x_k$ on the detector plane, and $N=\sum_k N(x_k)$. 

\begin{figure*}[t]
\centering
\includegraphics[width=\textwidth]{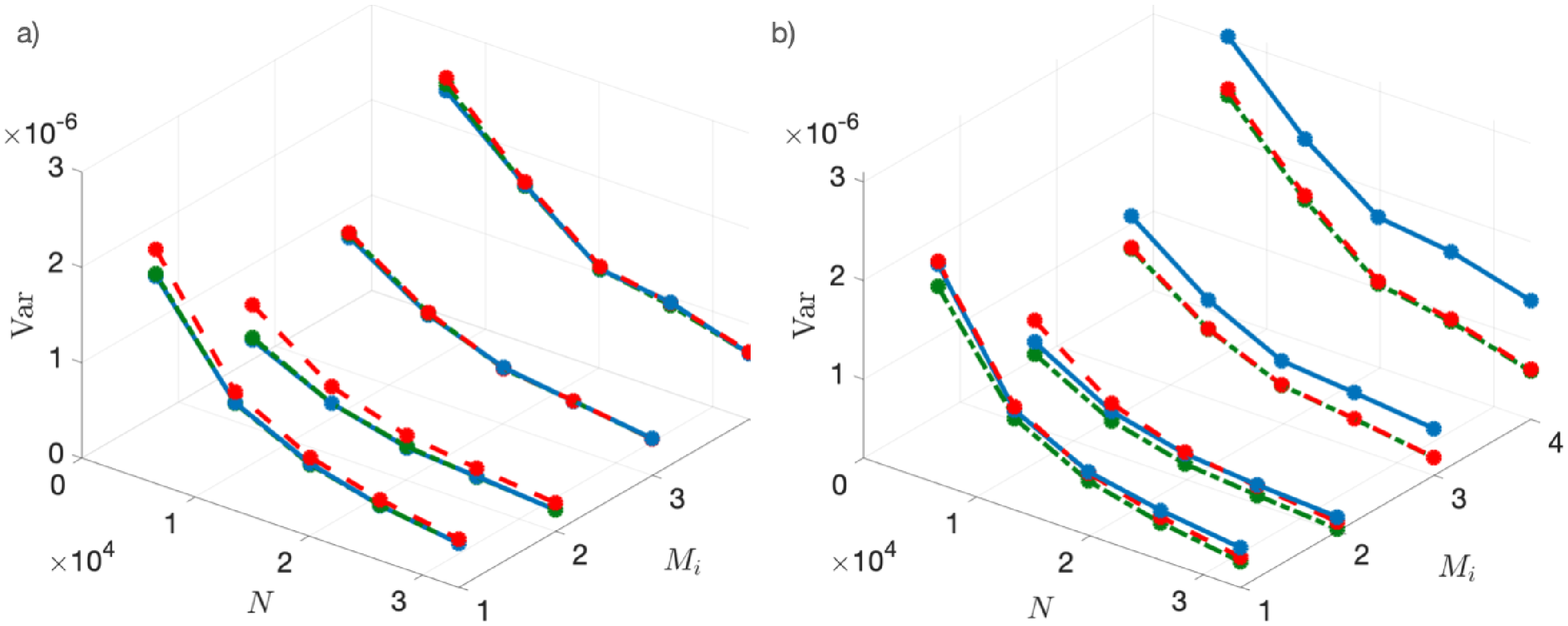}
\caption{a) Variances on the moments $M_i$ estimated through a Monte Carlo routine (4000 events) with Poissonian noise on $f(x)$ (blue circles, solid line) unconstrained Cramér-Rao Bound (red circles, dashed line) and constrained Cramér-Rao Bound (green circles, solid line)  b) Variances on the moments $M_i$ estimated through a Monte Carlo routine (4000 events) with Poissonian noise on the $f(x)$ and $H(x)$.}
\label{fig:results}
\end{figure*}

In our semiparametric approach, we deal with models with an infinity of degrees of freedom. In short, we want to estimate a small subset consisting of a finite number of parameters (in our case, moments), treating the others as a nuisance~\cite{Suzuki:2020aa}. Let the parameter of interest be an arbitrary linear combination of the moments:
\begin{equation}
    \label{eq:beta}
    \beta= \pmb{\mu}^\top \mathbf{M}=\pmb{\mu}^\top \mathbf{C}^{-1}\mathbf{m}\, ,
\end{equation}
where the superscript $\top$ denotes the transpose and the vector  $\pmb{\mu}$ can be arbitrarily chosen but we assume its components are independent of the $M_i$s. Its value can be estimated by direct data processing, leading to an estimator $\check{\beta}=\pmb{\mu}^\top \mathbf{C}^{-1} \check{\mathbf{m}}$.

To assess the performance of this estimator we are demanded to compare it to the expected Cram\'er-Rao bound from the Fisher information matrix. Each moment $M_i$ only depends on a finite number of moments $m_j$ with $j\leq i$~\cite{Tsang:2019aa}, so one can construct an estimator for each $M_i$ using the sample means of the measured moments $m_j$ and the finite number of instrumental response function moments that make up the first $i$ rows and columns of $\mathbf{C}$.  In this form, the problem might then superficially appear as a direct linear inversion, for which a variety of methods are available, the most prominent of which is perhaps the least squares~\cite{Lawson:1974aa}. However, this can mask pitfalls for the unwary: associating errors to this estimator is perilous, as the Fisher information matrix associated with any such estimator depends on all of the moments of the lineshape. Inverting this infinite-dimensional matrix and treating the infinite number of parameters other than $M_i$ as nuisance parameters to find a lower bound for precision in an estimate of $M_i$ is seldom possible. This is precisely the province of the semiparametric estimation~\cite{Bickel:1993aa,Tsiatis:2006aa} and its recent quantum version~\cite{Tsang:2020aa}, thus motivating our approach.

In the semiparametric scenario the Cram\'er-Rao bound can be established by means of an \emph{influence function}~\cite{Bickel:1993aa,Tsiatis:2006aa} that, for our parameter $\beta$, turns out to be $I_\beta(x) = \sum_{ij} \mu_i(\mathbf{C}^{-1})_{ij} x^j/N$. We normalise the spectrum to the total measured intensity $\int f(x)dx=N$ and the instrumental function to one $\int H(x)dx=1$: this amounts to considering the information per detected photon (see Supplementary Information). 
The influence function allows one to determine the Cram\'er-Rao bound as:
\begin{equation}
\label{eq:semiparaCRB}
        Var (\check{\beta}) \geq  \int I_\beta^2(x) f(x)  dx,
\end{equation}
where $Var ( \cdot ) $ denotes the variance. Here, $f(x)$ is then interpreted as the density pertaining to the statistical mean of a Poisson-distributed random variable $n(x)$: $d\bar{n}(x)=f(x)dx$. This lower bound is equivalent to the variance of the random functional $\int I_\beta(x) dn(x)$ (see Supplementary Information for more details). This ensures that the estimator \eqref{estimator} does achieve the ultimate limit when the sampling of $f(x)$ is sufficiently dense. 

In spectroscopy, the parameters of interest are often the moments with respect to a normalized object spectral line. This is equivalent to imposing $M_0=1$. Therefore, all the estimators for the moments will be defined as $\check{\beta}_{0}=\check{\beta}/\check{M}_0$. This new parameter will present a reduced variance, which can be bounded with the same semiparametric methods as \eqref{eq:semiparaCRB}
\begin{equation}
\label{eq:semiparaCRBc}
        Var(\check{\beta}_0) \geq  \int I_{{\rm eff},\beta}^2(x) f(x)  dx \, ,
\end{equation}
with a modified influence function $I_{{\rm eff},\beta}(x)=I_\beta(x)-\beta/N$. This leads to a lowering of the minimum by $\beta^2/N$, showing that constraint information is most important when one has employed fewer resources and less information is available.


In our experiment, we collected profiles, thus the coincidence numbers $N(x_k)$, for different counting times, resulting in a varying overall number of coincidences $N$. These are processed using the estimators \eqref{estimator}, then manipulated by means of \eqref{eq:beta}. As an example, we consider the estimation of the moments $M_1, M_2, M_3$ and $M_4$, with $M_0$ normalized to unity. The results in Fig.~\ref{fig:results} summarise the quantitative aspects of the achieved precision. Uncertainties are evaluated from a Monte Carlo routine generating Poissonian distributed outcomes, bootstrapping on the measured values. Figure~\ref{fig:results}a only considers the effect of the normalisation $M_0=1$ on the uncertainties. The red dots represent the expected unconstrained variance and the green dots the constrained one. These are evaluated by discretising the integrals in \eqref{eq:semiparaCRB} and \eqref{eq:semiparaCRBc}. The blue points refer to the Monte Carlo errors and, as expected for Poisson variables, sit exactly on the green dots. The effect of the constraint on the precision is more pronounced for $M_1$ and $M_2$, while it is reduced for the higher-order moments.

Crucially, the estimation can proceed from \eqref{eq:conv}, only after reconstructing the instrumental function $H(x)$. The analysis above, however, assumes that $H(x)$ is known with arbitrary precision. In fact, this is extracted from experimental data that carry their own uncertainty: quantifying how much these affect the final estimation is one of the main goals of this investigation. We have thus repeated our Monte Carlo routine to account for the experimental uncertainties on the evaluation of $H(x)$, and, consequently, $\mathbf{C}$. As above, we take Poisson variables. We notice that this is typically collected at much lower brightness since a narrower feature is required for a pseudo-pointlike source. The results are reported in Fig.~\ref{fig:results}b for the same four moments as before: the effect depends heavily on the order and, albeit modest for $M_1$ and $M_2$, it becomes more severe for $M_3$ and $M_4$. While the position and the width of the lines are reliably assessed, gathering information over more detail on the structure comes at a higher price. The main benefit of the semiparametric approach is ensuring that these precisions account for nuisance elements and are thus reliable.

In conclusion, the semiparametric theory set forth solves an important problem: the evaluation of the ultimate bounds on the attainable precision in the  deconvolution of lineshapes and the efficient estimation of spectral parameters when little prior information about the spectrum is available. We expect that these advanced statistical methods will bring further insights into many spectroscopic applications.

{\bf Funding} Horizon 2020 FET-OPEN-RIA STORMYTUNE (Grant No. 899587); Ministerio de Ciencia e Innovaci\'on (Grant PID2021-127781NB-I00); Natural Sciences and Engineering Research Council of Canada PDF program; National Research Council Canada's Quantum Sensors Challenge program.

{\bf Acknowledgments} We acknowledge helpful discussions with Andrea Chiuri, Fabio La Franca, Francesco Albarelli, and Vincenzo Berardi. AZG acknowledges that the NRC headquarters is located on the traditional unceded territory of the Algonquin Anishinaabe and Mohawk people.

{\bf Disclosures} The authors declare no conflicts of interest.\\

{\bf Data availability} Data underlying the results presented in this paper are not publicly available at this time but may be obtained from the authors upon reasonable request.\\

\section*{Appendix}

\subsection{Semiparametric estimation }
We summarise the treatment of semiparametric estimation originally presented in~\cite{Tsang:2019aa}. The response of the detector is described by means of a continuous set of outcomes $x$, which is abstractly called the detector space and may correspond to physical coordinate space. A distribution $dn(x)$ is defined in such a way that, for any subset $A$ of the detector space, one can measure the count rate $n(A)=\int_A dn(x)$. The value of $dn(x)$ is treated as a stochastic variable with mean $d\bar n(x)$ and Poisson distribution - implying that  finite count rates follow the same statistics, too. Given any vectorial quantity $\mathbf{h}(x)$ on the detector space, the expectation value can be written as the average over the mean counts:
\begin{equation}
    \mathrm{E}(\mathbf{h}) =\int \mathbf{h}(x)d\bar n(x),
\end{equation}
while the associated covariance matrix is 
\begin{equation}
    Var (\mathbf{h}) =\int \mathbf{h}(x)\mathbf{h}(x)^\top d\bar n(x).
\end{equation}

We can introduce a measure $d\mu(x)$ over the detector space, resulting in a density for the mean counts $f(x|\boldsymbol{\theta})=d\bar n(x)/d\mu(x)$, the so-called Radon-Nikodym derivative. This can depend on a vector $\boldsymbol{\theta}$ of parameters. This, in turn, sets a scalar parameter $\beta(\boldsymbol{\theta})$, representing our target. In principle, it is possible to establish a lower limit $CRB_{\beta}$ to the variance of any unbiased estimator $\check{\beta}$ by means of the Cram\'er-Rao bound (CRB). In the practice, however, a direct approach would demand the inversion of a large-dimension Fisher information matrix associated associated with the elements of $\boldsymbol{\theta}$, which may be infinite in number. The power of the semiparametric approach resides in establishing the CRB without resorting to inversion: indeed, this is obtained by a so-called influence function $I_{\beta}(x)$ associated to $\beta$ as
\begin{equation}
\label{eq:CRB}
\mathrm{CRB}_{\beta} = \int I_\beta(x)^2d\bar n(x).
\end{equation}
An influence function is readily found if one can write:
\begin{equation}
    \beta = \int I_\beta(x)f(x)d\mu(x),
\end{equation}
as it is the case in our experiment. The semiparametric treatment also prescribes how to modify the influence function to account for constraints on the parameters. These can indeed contribute to lowering the associated uncertainty, and this must be reflected in the CRB.

\subsection{Instrumental function for raster-scan imaging and spectrometry}
Consider a source with 1D intensity distribution $F(y)$ reaching a detection plane through an optical system with response function $S(x-y)$. If the contributions are incoherent, this results in a distribution $\tilde f(x)=\int S(x-y)F(y)dy$. This would be the quantity measured with a direct imaging detector at infinite resolution.

Instead, a raster scan on the detection plane gives a signal collected over a range $2\epsilon$:
\begin{equation}
\begin{aligned}
f(x)&=\int_{x-\epsilon}^{x+\epsilon} \tilde{f}(\xi)d\xi\\
&=\int_{x-\epsilon}^{x+\epsilon}\int S(\xi-y)F(y)d\xi dy
\end{aligned}
\end{equation}
Since there exists a function $\Sigma(\xi)$ such that $\int_{\xi_1}^{\xi_0 }S(\xi)d\xi=\Sigma(\xi_1)-\Sigma(\xi_0)$, we get
\begin{equation}
\begin{aligned}
\label{eq:direct}
f(x)&=\int \left(\Sigma(x-y+\epsilon)-\Sigma(x-y-\epsilon)\right)F(y)dy\\
&=\int H(x-y)F(y)dy.
\end{aligned}
\end{equation}
We thus obtain an instrumental function $H(x)$ for the raster scan, as in the case of simple direct imaging.

\subsection{Estimation of moments}
The profile $F(y)$ of the source can be normalised as $\int F(y)dy=1$, thus fixing the zero-order moment to $M_0=1$. The instrumental function is normalised to a factor $\tau$ giving the overall transmission of the system: $\int H(x)dx=\tau$. The detection function $f(x)$ is thus conveniently normalised as $\int f(x)dx=\tau N$ with $N$ being the total number of registered events. By setting $\tau=1$ in our investigation, we then consider the information per detected event.

The relation \eqref{eq:direct} between the real object $F(x)$ and the measured image $f(x)$ suggests evaluating moments by inversion
\begin{equation}
M_i =\sum_j (C^{-1})_{ij}m_j
\end{equation}
as explained in the main text. These relations help us in introducing proper influence functions for the estimators
\begin{equation}
I_i(x)= \frac{1}{N}\sum_j (C^{-1})_{ij}x^j,
\end{equation}
each associated to $M_i$. However, a CRB simply obtained as in~\eqref{eq:CRB} would neglect the effect of constraining $M_0=1$. This demands a modified influence function $I_{{\rm eff},i}(x)$ that the semiparametric estimation theory provides as
\begin{equation}
I_{{\rm eff},i}(x)=I_i(x)-\frac{M_i}{\tau N}.
\end{equation}

\subsection{Effect of the finite resolution on the Kramers-Kronig relations}
The spectral phase $\varphi(\omega)$ imparted by a sample can be retrieved by measurements of its spectral transmission $\eta(\omega)$ by means of the Kramers-Kronig relations
\begin{equation}
\label{eq:KKphase}
\varphi(\omega)=-\frac{1}{2\pi}\int \frac{\log(\eta(\omega))}{\omega'-\omega}d\omega.
\end{equation}
The transmission is ideally obtained as the ratio between a measured intensity $F_\eta(\omega)$ and a reference intensity $F_{\rm ref}(\omega)$:
\begin{equation}
\eta(\omega)=\frac{F_\eta(\omega)}{F_{\rm ref}(\omega)}.
\end{equation}
This is actually affected by the instrumental function $H(\omega)$ as
\begin{equation}
\eta_H(\omega)=\frac{\int H(\omega-\omega')F_\eta(\omega')}{\int H(\omega-\omega')F_{\rm ref}(\omega')}.
\end{equation}
The deviation of the exact expression for the phase \eqref{eq:KKphase} with the one obtained using $\eta_H$ is then
\begin{equation}
\label{eq:discrepancy}
    \Delta\varphi(\omega)=-\frac{1}{2\pi}\int\frac{1}{\omega'-\omega} \log\left(\frac{\int H(\omega'-\omega")F_\eta(\omega")}{F_\eta(\omega')}\right)d\omega',
\end{equation}
{\it i.e.} it reads as the Hilbert transform of a suitable function. In this expression \eqref{eq:discrepancy} we have made the assumption that the reference intensity $F_{\rm ref}$ varies slowly over a much wider interval than the width of $H(\omega)$, so we can neglect its error. By the properties of the Hilbert transform, we can thus obtain the quadratic error
\begin{equation}
\begin{aligned}
    &\epsilon^2 = \int \Delta\varphi(\omega)^2d\omega\\
    &= \frac{1}{4}\int \log^2\left(\frac{\int H(\omega-\omega')F_\eta(\omega')}{F_\eta(\omega)}\right)d\omega \, .
\end{aligned}
\end{equation}
\bigskip

\bibliography{PSF}

\begin{thebibliography}{30}%
\makeatletter
\providecommand \@ifxundefined [1]{%
 \@ifx{#1\undefined}
}%
\providecommand \@ifnum [1]{%
 \ifnum #1\expandafter \@firstoftwo
 \else \expandafter \@secondoftwo
 \fi
}%
\providecommand \@ifx [1]{%
 \ifx #1\expandafter \@firstoftwo
 \else \expandafter \@secondoftwo
 \fi
}%
\providecommand \natexlab [1]{#1}%
\providecommand \enquote  [1]{``#1''}%
\providecommand \bibnamefont  [1]{#1}%
\providecommand \bibfnamefont [1]{#1}%
\providecommand \citenamefont [1]{#1}%
\providecommand \href@noop [0]{\@secondoftwo}%
\providecommand \href [0]{\begingroup \@sanitize@url \@href}%
\providecommand \@href[1]{\@@startlink{#1}\@@href}%
\providecommand \@@href[1]{\endgroup#1\@@endlink}%
\providecommand \@sanitize@url [0]{\catcode `\\12\catcode `\$12\catcode
  `\&12\catcode `\#12\catcode `\^12\catcode `\_12\catcode `\%12\relax}%
\providecommand \@@startlink[1]{}%
\providecommand \@@endlink[0]{}%
\providecommand \url  [0]{\begingroup\@sanitize@url \@url }%
\providecommand \@url [1]{\endgroup\@href {#1}{\urlprefix }}%
\providecommand \urlprefix  [0]{URL }%
\providecommand \Eprint [0]{\href }%
\providecommand \doibase [0]{https://doi.org/}%
\providecommand \selectlanguage [0]{\@gobble}%
\providecommand \bibinfo  [0]{\@secondoftwo}%
\providecommand \bibfield  [0]{\@secondoftwo}%
\providecommand \translation [1]{[#1]}%
\providecommand \BibitemOpen [0]{}%
\providecommand \bibitemStop [0]{}%
\providecommand \bibitemNoStop [0]{.\EOS\space}%
\providecommand \EOS [0]{\spacefactor3000\relax}%
\providecommand \BibitemShut  [1]{\csname bibitem#1\endcsname}%
\let\auto@bib@innerbib\@empty
\bibitem [{\citenamefont {Rinnan}\ \emph {et~al.}(2009)\citenamefont {Rinnan},
  \citenamefont {Berg},\ and\ \citenamefont {Engelsen}}]{Rinnan:2009aa}%
  \BibitemOpen
  \bibfield  {author} {\bibinfo {author} {\bibfnamefont {{\AA}.}~\bibnamefont
  {Rinnan}}, \bibinfo {author} {\bibfnamefont {F.~v.~d.}\ \bibnamefont
  {Berg}},\ and\ \bibinfo {author} {\bibfnamefont {S.~B.}\ \bibnamefont
  {Engelsen}},\ }\bibfield  {title} {\bibinfo {title} {Review of the most
  common pre-processing techniques for near-infrared spectra},\ }\href
  {https://doi.org/https://doi.org/10.1016/j.trac.2009.07.007} {\bibfield
  {journal} {\bibinfo  {journal} {TaAC Trend. Anal. Chem.}\ }\textbf {\bibinfo
  {volume} {28}},\ \bibinfo {pages} {1201} (\bibinfo {year}
  {2009})}\BibitemShut {NoStop}%
\bibitem [{\citenamefont {Dubrovkin}(2021)}]{Dubrovkin:2021aa}%
  \BibitemOpen
  \bibfield  {author} {\bibinfo {author} {\bibfnamefont {J.}~\bibnamefont
  {Dubrovkin}},\ }\href@noop {} {\emph {\bibinfo {title} {Derivative
  Spectroscopy}}}\ (\bibinfo  {publisher} {Cambridge Scholars Publishing},\
  \bibinfo {year} {2021})\BibitemShut {NoStop}%
\bibitem [{\citenamefont {Kauppinen}\ \emph {et~al.}(1981)\citenamefont
  {Kauppinen}, \citenamefont {Moffatt}, \citenamefont {Mantsch},\ and\
  \citenamefont {Cameron}}]{Kauppinen:1981aa}%
  \BibitemOpen
  \bibfield  {author} {\bibinfo {author} {\bibfnamefont {J.~K.}\ \bibnamefont
  {Kauppinen}}, \bibinfo {author} {\bibfnamefont {D.~J.}\ \bibnamefont
  {Moffatt}}, \bibinfo {author} {\bibfnamefont {H.~H.}\ \bibnamefont
  {Mantsch}},\ and\ \bibinfo {author} {\bibfnamefont {D.~G.}\ \bibnamefont
  {Cameron}},\ }\bibfield  {title} {\bibinfo {title} {Fourier
  self-deconvolution: A method for resolving intrinsically overlapped bands},\
  }\href {https://opg.optica.org/as/abstract.cfm?URI=as-35-3-271} {\bibfield
  {journal} {\bibinfo  {journal} {Appl. Spectrosc.}\ }\textbf {\bibinfo
  {volume} {35}},\ \bibinfo {pages} {271} (\bibinfo {year} {1981})}\BibitemShut
  {NoStop}%
\bibitem [{\citenamefont {Gelb}(1974)}]{Gelb:1974aa}%
  \BibitemOpen
  \bibfield  {author} {\bibinfo {author} {\bibfnamefont {A.}~\bibnamefont
  {Gelb}},\ }\href@noop {} {\emph {\bibinfo {title} {Applied optimal
  estimation}}}\ (\bibinfo  {publisher} {MIT Press},\ \bibinfo {year}
  {1974})\BibitemShut {NoStop}%
\bibitem [{\citenamefont {Sprzeczak}\ and\ \citenamefont
  {Morawski}(2001)}]{Sprzeczak:2001aa}%
  \BibitemOpen
  \bibfield  {author} {\bibinfo {author} {\bibfnamefont {P.}~\bibnamefont
  {Sprzeczak}}\ and\ \bibinfo {author} {\bibfnamefont {R.~Z.}\ \bibnamefont
  {Morawski}},\ }\bibfield  {title} {\bibinfo {title} {Cauchy-filter-based
  algorithms for reconstruction of absorption spectra},\ }\href
  {https://doi.org/10.1109/19.963170} {\bibfield  {journal} {\bibinfo
  {journal} {IEEE Trans. Instrum. Meas.}\ }\textbf {\bibinfo {volume} {50}},\
  \bibinfo {pages} {1123} (\bibinfo {year} {2001})}\BibitemShut {NoStop}%
\bibitem [{\citenamefont {Averbuch}\ and\ \citenamefont
  {Zheludev}(2009)}]{Averbuch:2009aa}%
  \BibitemOpen
  \bibfield  {author} {\bibinfo {author} {\bibfnamefont {A.}~\bibnamefont
  {Averbuch}}\ and\ \bibinfo {author} {\bibfnamefont {V.}~\bibnamefont
  {Zheludev}},\ }\bibfield  {title} {\bibinfo {title} {Spline-based
  deconvolution},\ }\href
  {https://doi.org/https://doi.org/10.1016/j.sigpro.2009.03.022} {\bibfield
  {journal} {\bibinfo  {journal} {Signal Process.}\ }\textbf {\bibinfo {volume}
  {89}},\ \bibinfo {pages} {1782} (\bibinfo {year} {2009})}\BibitemShut
  {NoStop}%
\bibitem [{\citenamefont {Liu}\ \emph {et~al.}(2013)\citenamefont {Liu},
  \citenamefont {Yan}, \citenamefont {Chang}, \citenamefont {Fang},\ and\
  \citenamefont {Zhang}}]{Liu:2013aa}%
  \BibitemOpen
  \bibfield  {author} {\bibinfo {author} {\bibfnamefont {H.}~\bibnamefont
  {Liu}}, \bibinfo {author} {\bibfnamefont {L.}~\bibnamefont {Yan}}, \bibinfo
  {author} {\bibfnamefont {Y.}~\bibnamefont {Chang}}, \bibinfo {author}
  {\bibfnamefont {H.}~\bibnamefont {Fang}},\ and\ \bibinfo {author}
  {\bibfnamefont {T.}~\bibnamefont {Zhang}},\ }\bibfield  {title} {\bibinfo
  {title} {Spectral deconvolution and feature extraction with robust adaptive
  tikhonov regularization},\ }\href {https://doi.org/10.1109/TIM.2012.2217636}
  {\bibfield  {journal} {\bibinfo  {journal} {IEEE Trans. Instrum. Meas.}\
  }\textbf {\bibinfo {volume} {62}},\ \bibinfo {pages} {315} (\bibinfo {year}
  {2013})}\BibitemShut {NoStop}%
\bibitem [{\citenamefont {Navab}\ \emph {et~al.}(2015)\citenamefont {Navab},
  \citenamefont {Hornegger}, \citenamefont {Wells},\ and\ \citenamefont
  {Frangi}}]{Ronneberger:2015aa}%
  \BibitemOpen
  \bibinfo {editor} {\bibfnamefont {N.}~\bibnamefont {Navab}}, \bibinfo
  {editor} {\bibfnamefont {J.}~\bibnamefont {Hornegger}}, \bibinfo {editor}
  {\bibfnamefont {W.~M.}\ \bibnamefont {Wells}},\ and\ \bibinfo {editor}
  {\bibfnamefont {A.~F.}\ \bibnamefont {Frangi}},\ eds.,\ \href@noop {} {\emph
  {\bibinfo {title} {Medical Image Computing and Computer-Assisted Intervention
  --MICCAI 2015}}}\ (\bibinfo  {publisher} {Springer},\ \bibinfo {address}
  {Cham},\ \bibinfo {year} {2015})\BibitemShut {NoStop}%
\bibitem [{\citenamefont {Yanny}\ \emph {et~al.}(2022)\citenamefont {Yanny},
  \citenamefont {Monakhova}, \citenamefont {Shuai},\ and\ \citenamefont
  {Waller}}]{Yanny:2022aa}%
  \BibitemOpen
  \bibfield  {author} {\bibinfo {author} {\bibfnamefont {K.}~\bibnamefont
  {Yanny}}, \bibinfo {author} {\bibfnamefont {K.}~\bibnamefont {Monakhova}},
  \bibinfo {author} {\bibfnamefont {R.~W.}\ \bibnamefont {Shuai}},\ and\
  \bibinfo {author} {\bibfnamefont {L.}~\bibnamefont {Waller}},\ }\bibfield
  {title} {\bibinfo {title} {Deep learning for fast spatially varying
  deconvolution},\ }\href {https://doi.org/10.1364/OPTICA.442438} {\bibfield
  {journal} {\bibinfo  {journal} {Optica}\ }\textbf {\bibinfo {volume} {9}},\
  \bibinfo {pages} {96} (\bibinfo {year} {2022})}\BibitemShut {NoStop}%
\bibitem [{\citenamefont {Torrisi}\ \emph {et~al.}(2020)\citenamefont
  {Torrisi}, \citenamefont {Carbone}, \citenamefont {Rohr}, \citenamefont
  {Montoya}, \citenamefont {Ha}, \citenamefont {Yano}, \citenamefont {Suram},\
  and\ \citenamefont {Hung}}]{Torrisi:2020aa}%
  \BibitemOpen
  \bibfield  {author} {\bibinfo {author} {\bibfnamefont {S.~B.}\ \bibnamefont
  {Torrisi}}, \bibinfo {author} {\bibfnamefont {M.~R.}\ \bibnamefont
  {Carbone}}, \bibinfo {author} {\bibfnamefont {B.~A.}\ \bibnamefont {Rohr}},
  \bibinfo {author} {\bibfnamefont {J.~H.}\ \bibnamefont {Montoya}}, \bibinfo
  {author} {\bibfnamefont {Y.}~\bibnamefont {Ha}}, \bibinfo {author}
  {\bibfnamefont {J.}~\bibnamefont {Yano}}, \bibinfo {author} {\bibfnamefont
  {S.~K.}\ \bibnamefont {Suram}},\ and\ \bibinfo {author} {\bibfnamefont
  {L.}~\bibnamefont {Hung}},\ }\bibfield  {title} {\bibinfo {title} {Random
  forest machine learning models for interpretable x-ray absorption near-edge
  structure spectrum-property relationships},\ }\href
  {https://doi.org/10.1038/s41524-020-00376-6} {\bibfield  {journal} {\bibinfo
  {journal} {npj Comput. Mater.}\ }\textbf {\bibinfo {volume} {6}},\ \bibinfo
  {pages} {109} (\bibinfo {year} {2020})}\BibitemShut {NoStop}%
\bibitem [{\citenamefont {Albarelli}\ \emph {et~al.}(2020)\citenamefont
  {Albarelli}, \citenamefont {Barbieri}, \citenamefont {Genoni},\ and\
  \citenamefont {Gianani}}]{Albarelli:2020aa}%
  \BibitemOpen
  \bibfield  {author} {\bibinfo {author} {\bibfnamefont {F.}~\bibnamefont
  {Albarelli}}, \bibinfo {author} {\bibfnamefont {M.}~\bibnamefont {Barbieri}},
  \bibinfo {author} {\bibfnamefont {M.~G.}\ \bibnamefont {Genoni}},\ and\
  \bibinfo {author} {\bibfnamefont {I.}~\bibnamefont {Gianani}},\ }\bibfield
  {title} {\bibinfo {title} {A perspective on multiparameter quantum metrology:
  From theoretical tools to applications in quantum imaging},\ }\href
  {https://doi.org/https://doi.org/10.1016/j.physleta.2020.126311} {\bibfield
  {journal} {\bibinfo  {journal} {Phys. Lett. A}\ }\textbf {\bibinfo {volume}
  {384}},\ \bibinfo {pages} {126311} (\bibinfo {year} {2020})}\BibitemShut
  {NoStop}%
\bibitem [{\citenamefont {Polino}\ \emph {et~al.}(2020)\citenamefont {Polino},
  \citenamefont {Valeri}, \citenamefont {Spagnolo},\ and\ \citenamefont
  {Sciarrino}}]{Polino:2020aa}%
  \BibitemOpen
  \bibfield  {author} {\bibinfo {author} {\bibfnamefont {E.}~\bibnamefont
  {Polino}}, \bibinfo {author} {\bibfnamefont {M.}~\bibnamefont {Valeri}},
  \bibinfo {author} {\bibfnamefont {N.}~\bibnamefont {Spagnolo}},\ and\
  \bibinfo {author} {\bibfnamefont {F.}~\bibnamefont {Sciarrino}},\ }\bibfield
  {title} {\bibinfo {title} {Photonic quantum metrology},\ }\href
  {https://doi.org/10.1116/5.0007577} {\bibfield  {journal} {\bibinfo
  {journal} {AVS Quantum Sci.}\ }\textbf {\bibinfo {volume} {2}},\ \bibinfo
  {pages} {024703} (\bibinfo {year} {2020})}\BibitemShut {NoStop}%
\bibitem [{\citenamefont {Amiot}\ \emph {et~al.}(2018)\citenamefont {Amiot},
  \citenamefont {Ryczkowski}, \citenamefont {Friberg}, \citenamefont {Dudley},\
  and\ \citenamefont {Genty}}]{Amiot:2018aa}%
  \BibitemOpen
  \bibfield  {author} {\bibinfo {author} {\bibfnamefont {C.}~\bibnamefont
  {Amiot}}, \bibinfo {author} {\bibfnamefont {P.}~\bibnamefont {Ryczkowski}},
  \bibinfo {author} {\bibfnamefont {A.~T.}\ \bibnamefont {Friberg}}, \bibinfo
  {author} {\bibfnamefont {J.~M.}\ \bibnamefont {Dudley}},\ and\ \bibinfo
  {author} {\bibfnamefont {G.}~\bibnamefont {Genty}},\ }\bibfield  {title}
  {\bibinfo {title} {Supercontinuum spectral-domain ghost imaging},\ }\href
  {https://doi.org/10.1364/OL.43.005025} {\bibfield  {journal} {\bibinfo
  {journal} {Opt. Lett.}\ }\textbf {\bibinfo {volume} {43}},\ \bibinfo {pages}
  {5025} (\bibinfo {year} {2018})}\BibitemShut {NoStop}%
\bibitem [{\citenamefont {Janassek}\ \emph {et~al.}(2018)\citenamefont
  {Janassek}, \citenamefont {Herdt}, \citenamefont {Blumenstein},\ and\
  \citenamefont {Els{\"a}sser}}]{Janassek:2018aa}%
  \BibitemOpen
  \bibfield  {author} {\bibinfo {author} {\bibfnamefont {P.}~\bibnamefont
  {Janassek}}, \bibinfo {author} {\bibfnamefont {A.}~\bibnamefont {Herdt}},
  \bibinfo {author} {\bibfnamefont {S.}~\bibnamefont {Blumenstein}},\ and\
  \bibinfo {author} {\bibfnamefont {W.}~\bibnamefont {Els{\"a}sser}},\
  }\bibfield  {title} {\bibinfo {title} {Ghost spectroscopy with classical
  correlated amplified spontaneous emission photons emitted by an erbium-doped
  fiber amplifier},\ }\bibfield  {journal} {\bibinfo  {journal} {Appl. Sci.}\
  }\textbf {\bibinfo {volume} {8}},\ \href {https://doi.org/10.3390/app8101896}
  {10.3390/app8101896} (\bibinfo {year} {2018})\BibitemShut {NoStop}%
\bibitem [{\citenamefont {Chiuri}\ \emph {et~al.}(2022)\citenamefont {Chiuri},
  \citenamefont {Gianani}, \citenamefont {Cimini}, \citenamefont
  {De~Dominicis}, \citenamefont {Genoni},\ and\ \citenamefont
  {Barbieri}}]{Chiuri:2022aa}%
  \BibitemOpen
  \bibfield  {author} {\bibinfo {author} {\bibfnamefont {A.}~\bibnamefont
  {Chiuri}}, \bibinfo {author} {\bibfnamefont {I.}~\bibnamefont {Gianani}},
  \bibinfo {author} {\bibfnamefont {V.}~\bibnamefont {Cimini}}, \bibinfo
  {author} {\bibfnamefont {L.}~\bibnamefont {De~Dominicis}}, \bibinfo {author}
  {\bibfnamefont {M.~G.}\ \bibnamefont {Genoni}},\ and\ \bibinfo {author}
  {\bibfnamefont {M.}~\bibnamefont {Barbieri}},\ }\bibfield  {title} {\bibinfo
  {title} {Ghost imaging as loss estimation: Quantum versus classical
  schemes},\ }\href {https://doi.org/10.1103/PhysRevA.105.013506} {\bibfield
  {journal} {\bibinfo  {journal} {Phys. Rev. A}\ }\textbf {\bibinfo {volume}
  {105}},\ \bibinfo {pages} {013506} (\bibinfo {year} {2022})}\BibitemShut
  {NoStop}%
\bibitem [{\citenamefont {Tsang}(2019)}]{Tsang:2019aa}%
  \BibitemOpen
  \bibfield  {author} {\bibinfo {author} {\bibfnamefont {M.}~\bibnamefont
  {Tsang}},\ }\bibfield  {title} {\bibinfo {title} {Semiparametric estimation
  for incoherent optical imaging},\ }\href
  {https://doi.org/10.1103/PhysRevResearch.1.033006} {\bibfield  {journal}
  {\bibinfo  {journal} {Phys. Rev. Research}\ }\textbf {\bibinfo {volume}
  {1}},\ \bibinfo {pages} {033006} (\bibinfo {year} {2019})}\BibitemShut
  {NoStop}%
\bibitem [{\citenamefont {Cimini}\ \emph {et~al.}(2021)\citenamefont {Cimini},
  \citenamefont {Albarelli}, \citenamefont {Gianani},\ and\ \citenamefont
  {Barbieri}}]{Cimini:2021aa}%
  \BibitemOpen
  \bibfield  {author} {\bibinfo {author} {\bibfnamefont {V.}~\bibnamefont
  {Cimini}}, \bibinfo {author} {\bibfnamefont {F.}~\bibnamefont {Albarelli}},
  \bibinfo {author} {\bibfnamefont {I.}~\bibnamefont {Gianani}},\ and\ \bibinfo
  {author} {\bibfnamefont {M.}~\bibnamefont {Barbieri}},\ }\bibfield  {title}
  {\bibinfo {title} {Semiparametric estimation of the {Hong-Ou-Mandel}
  profile},\ }\href {https://doi.org/10.1103/PhysRevA.104.L061701} {\bibfield
  {journal} {\bibinfo  {journal} {Phys. Rev. A}\ }\textbf {\bibinfo {volume}
  {104}},\ \bibinfo {pages} {L061701} (\bibinfo {year} {2021})}\BibitemShut
  {NoStop}%
\bibitem [{\citenamefont {Pike}\ \emph {et~al.}(1984)\citenamefont {Pike},
  \citenamefont {McWhirter}, \citenamefont {Bertero},\ and\ \citenamefont
  {de~Mol}}]{Pike:1984aa}%
  \BibitemOpen
  \bibfield  {author} {\bibinfo {author} {\bibfnamefont {E.}~\bibnamefont
  {Pike}}, \bibinfo {author} {\bibfnamefont {J.}~\bibnamefont {McWhirter}},
  \bibinfo {author} {\bibfnamefont {M.}~\bibnamefont {Bertero}},\ and\ \bibinfo
  {author} {\bibfnamefont {C.}~\bibnamefont {de~Mol}},\ }\bibfield  {title}
  {\bibinfo {title} {Generalised information theory for inverse problems in
  signal processing},\ }in\ \href
  {https://digital-library.theiet.org/content/journals/10.1049/ip-f-1.1984.0100}
  {\emph {\bibinfo {booktitle} {IEE Proceedings}}},\ Vol.\ \bibinfo {volume}
  {131}\ (\bibinfo {year} {1984})\ pp.\ \bibinfo {pages} {660--667}\BibitemShut
  {NoStop}%
\bibitem [{\citenamefont {Rautian}(1958)}]{Rautian:1958aa}%
  \BibitemOpen
  \bibfield  {author} {\bibinfo {author} {\bibfnamefont {S.~G.}\ \bibnamefont
  {Rautian}},\ }\bibfield  {title} {\bibinfo {title} {Real spectral
  instruments},\ }\href@noop {} {\bibfield  {journal} {\bibinfo  {journal}
  {Usp. fiz. nauk}\ }\textbf {\bibinfo {volume} {46}},\ \bibinfo {pages} {475}
  (\bibinfo {year} {1958})}\BibitemShut {NoStop}%
\bibitem [{\citenamefont {Chen}\ \emph {et~al.}(2021)\citenamefont {Chen},
  \citenamefont {Treu}, \citenamefont {Fassnacht}, \citenamefont {Ragland},
  \citenamefont {Schmidt},\ and\ \citenamefont {Suyu}}]{Chen:2021aa}%
  \BibitemOpen
  \bibfield  {author} {\bibinfo {author} {\bibfnamefont {G.~C.-F.}\
  \bibnamefont {Chen}}, \bibinfo {author} {\bibfnamefont {T.}~\bibnamefont
  {Treu}}, \bibinfo {author} {\bibfnamefont {C.~D.}\ \bibnamefont {Fassnacht}},
  \bibinfo {author} {\bibfnamefont {S.}~\bibnamefont {Ragland}}, \bibinfo
  {author} {\bibfnamefont {T.}~\bibnamefont {Schmidt}},\ and\ \bibinfo {author}
  {\bibfnamefont {S.~H.}\ \bibnamefont {Suyu}},\ }\bibfield  {title} {\bibinfo
  {title} {Point spread function reconstruction of adaptive-optics imaging:
  meeting the astrometric requirements for time-delay cosmography},\ }\bibfield
   {booktitle} {\emph {\bibinfo {booktitle} {Monthly Notices of the Royal
  Astronomical Society}},\ }\href {https://doi.org/10.1093/mnras/stab2587}
  {\bibfield  {journal} {\bibinfo  {journal} {Mon. Notices Royal Astron. Soc.}\
  }\textbf {\bibinfo {volume} {508}},\ \bibinfo {pages} {755} (\bibinfo {year}
  {2021})}\BibitemShut {NoStop}%
\bibitem [{\citenamefont {Kay}(1993)}]{Kay:1993aa}%
  \BibitemOpen
  \bibfield  {author} {\bibinfo {author} {\bibfnamefont {S.~M.}\ \bibnamefont
  {Kay}},\ }\href@noop {} {\emph {\bibinfo {title} {Fundamentals of Statistical
  Processing: Estimation Theory}}}\ (\bibinfo  {publisher} {Prentice Hall},\
  \bibinfo {year} {1993})\BibitemShut {NoStop}%
\bibitem [{\citenamefont {Gordon}(1965)}]{Gordon:1965aa}%
  \BibitemOpen
  \bibfield  {author} {\bibinfo {author} {\bibfnamefont {R.~G.}\ \bibnamefont
  {Gordon}},\ }\bibfield  {title} {\bibinfo {title} {Molecular motion in
  infrared and {Raman} spectra},\ }\href {https://doi.org/10.1063/1.1696920}
  {\bibfield  {journal} {\bibinfo  {journal} {J. Chem. Phys.}\ }\textbf
  {\bibinfo {volume} {43}},\ \bibinfo {pages} {1307} (\bibinfo {year}
  {1965})}\BibitemShut {NoStop}%
\bibitem [{\citenamefont {Leine}(1968)}]{Leine:1968aa}%
  \BibitemOpen
  \bibfield  {author} {\bibinfo {author} {\bibfnamefont {H.~B.}\ \bibnamefont
  {Leine}},\ }\bibfield  {title} {\bibinfo {title} {Role of dispersion in
  collision-induced absorption},\ }\href
  {https://doi.org/10.1103/PhysRevLett.21.1512} {\bibfield  {journal} {\bibinfo
   {journal} {Phys. Rev. Lett.}\ }\textbf {\bibinfo {volume} {21}},\ \bibinfo
  {pages} {1512} (\bibinfo {year} {1968})}\BibitemShut {NoStop}%
\bibitem [{\citenamefont {Jacobson}(1971{\natexlab{a}})}]{Jacobson:1971aa}%
  \BibitemOpen
  \bibfield  {author} {\bibinfo {author} {\bibfnamefont {H.~C.}\ \bibnamefont
  {Jacobson}},\ }\bibfield  {title} {\bibinfo {title} {Moment analysis of
  atomic spectral lines},\ }\href {https://doi.org/10.1103/PhysRevA.4.1363}
  {\bibfield  {journal} {\bibinfo  {journal} {Phys. Rev. A}\ }\textbf {\bibinfo
  {volume} {4}},\ \bibinfo {pages} {1363} (\bibinfo {year}
  {1971}{\natexlab{a}})}\BibitemShut {NoStop}%
\bibitem [{\citenamefont {Jacobson}(1971{\natexlab{b}})}]{Jacobson:1971ab}%
  \BibitemOpen
  \bibfield  {author} {\bibinfo {author} {\bibfnamefont {H.~C.}\ \bibnamefont
  {Jacobson}},\ }\bibfield  {title} {\bibinfo {title} {Intermolecular forces
  from atomic line-shape experiments},\ }\href
  {https://doi.org/10.1103/PhysRevA.4.1368} {\bibfield  {journal} {\bibinfo
  {journal} {Phys. Rev. A}\ }\textbf {\bibinfo {volume} {4}},\ \bibinfo {pages}
  {1368} (\bibinfo {year} {1971}{\natexlab{b}})}\BibitemShut {NoStop}%
\bibitem [{\citenamefont {Suzuki}\ \emph {et~al.}(2020)\citenamefont {Suzuki},
  \citenamefont {Yang},\ and\ \citenamefont {Hayashi}}]{Suzuki:2020aa}%
  \BibitemOpen
  \bibfield  {author} {\bibinfo {author} {\bibfnamefont {J.}~\bibnamefont
  {Suzuki}}, \bibinfo {author} {\bibfnamefont {Y.}~\bibnamefont {Yang}},\ and\
  \bibinfo {author} {\bibfnamefont {M.}~\bibnamefont {Hayashi}},\ }\bibfield
  {title} {\bibinfo {title} {Quantum state estimation with nuisance
  parameters},\ }\href {https://doi.org/10.1088/1751-8121/ab8b78} {\bibfield
  {journal} {\bibinfo  {journal} {J. Phys. A: Math. Theor.}\ }\textbf {\bibinfo
  {volume} {53}},\ \bibinfo {pages} {453001} (\bibinfo {year}
  {2020})}\BibitemShut {NoStop}%
\bibitem [{\citenamefont {Lawson}\ and\ \citenamefont
  {Hanson}(1974)}]{Lawson:1974aa}%
  \BibitemOpen
  \bibfield  {author} {\bibinfo {author} {\bibfnamefont {C.}~\bibnamefont
  {Lawson}}\ and\ \bibinfo {author} {\bibfnamefont {R.}~\bibnamefont
  {Hanson}},\ }\href@noop {} {\emph {\bibinfo {title} {Solving Least Squares
  Problems}}}\ (\bibinfo  {publisher} {Prentice-Hall},\ \bibinfo {address}
  {Englewood Cliffs},\ \bibinfo {year} {1974})\BibitemShut {NoStop}%
\bibitem [{\citenamefont {Bickel}\ \emph {et~al.}(1993)\citenamefont {Bickel},
  \citenamefont {Klaassen}, \citenamefont {Ritov},\ and\ \citenamefont
  {Wellner}}]{Bickel:1993aa}%
  \BibitemOpen
  \bibfield  {author} {\bibinfo {author} {\bibfnamefont {P.~J.}\ \bibnamefont
  {Bickel}}, \bibinfo {author} {\bibfnamefont {C.~A.~J.}\ \bibnamefont
  {Klaassen}}, \bibinfo {author} {\bibfnamefont {Y.}~\bibnamefont {Ritov}},\
  and\ \bibinfo {author} {\bibfnamefont {J.~A.}\ \bibnamefont {Wellner}},\
  }\href@noop {} {\emph {\bibinfo {title} {Efficient and Adaptive Estimation
  for Semiparametric Models}}}\ (\bibinfo  {publisher} {Springer},\ \bibinfo
  {year} {1993})\BibitemShut {NoStop}%
\bibitem [{\citenamefont {Tsiatis}(2006)}]{Tsiatis:2006aa}%
  \BibitemOpen
  \bibfield  {author} {\bibinfo {author} {\bibfnamefont {A.}~\bibnamefont
  {Tsiatis}},\ }\href@noop {} {\emph {\bibinfo {title} {Semiparametric Theory
  and Missing Data}}}\ (\bibinfo  {publisher} {Springer},\ \bibinfo {year}
  {2006})\BibitemShut {NoStop}%
\bibitem [{\citenamefont {Tsang}\ \emph {et~al.}(2020)\citenamefont {Tsang},
  \citenamefont {Albarelli},\ and\ \citenamefont {Datta}}]{Tsang:2020aa}%
  \BibitemOpen
  \bibfield  {author} {\bibinfo {author} {\bibfnamefont {M.}~\bibnamefont
  {Tsang}}, \bibinfo {author} {\bibfnamefont {F.}~\bibnamefont {Albarelli}},\
  and\ \bibinfo {author} {\bibfnamefont {A.}~\bibnamefont {Datta}},\ }\bibfield
   {title} {\bibinfo {title} {Quantum semiparametric estimation},\ }\href
  {https://doi.org/10.1103/PhysRevX.10.031023} {\bibfield  {journal} {\bibinfo
  {journal} {Phys. Rev. X}\ }\textbf {\bibinfo {volume} {10}},\ \bibinfo
  {pages} {031023} (\bibinfo {year} {2020})}\BibitemShut {NoStop}%
\end{thebibliography}%

\end{document}